\documentclass[12pt]{article}

\usepackage{natbib}
\usepackage{amsmath}
\usepackage{subcaption}
\usepackage{graphicx}

\title{Diagnostics for Stochastic Gaussian Process Emulators}
\author{Evan Baker \thanks{Primary and corresponding author: Department of Mathematics, University of Exeter; \texttt{e.baker@exeter.ac.uk}; Laver Building, North Park Road, Exeter, EX4 4QF},
 Peter Challenor \thanks{Department of Mathematics, University of Exeter, \tt P.G.Challenor@exeter.ac.uk},
  and Matt Eames \thanks{Department of Engineering, University of Exeter \tt M.E.Eames@exeter.ac.uk}}
\date{}

\begin{document}

\maketitle

\begin{abstract}
Computer models, also known as simulators, can be computationally expensive to run, and for this reason statistical surrogates, known as emulators, are often used. Any statistical model, including an emulator, should be validated before being used, otherwise resulting decisions can be misguided. We discuss how current methods for validating Gaussian process emulators of deterministic models are insufficient for emulators of stochastic computer models and develop a framework for diagnosing problems in stochastic emulators. These diagnostics are based on independently validating the mean and variance predictions using out-of-sample, replicated, simulator runs. We then also use a building performance simulator as a case study example.
\end{abstract}


\section{Introduction}
\label{sec:Intro}

Increasingly, science relies on complex numerical models to aid in understanding and decision making. Climate science is a classic example of this, where incredibly complex computer models are used to learn key attributes about the climate system. Because of the computational cost of these models, faster statistical models are sometimes trained to act as a surrogate for the costly numerical models (the overuse of the word `model' often lead to the complex numerical model being called a `simulator' and the faster statistical model being called an `emulator'). Emulating complex computer models is a widely researched idea \citep{sacks1989design, kennedy2001bayesian, o2006bayesian, gramacy2020surrogates}, usually revolving around the use of a Gaussian process because of its flexibility, its comprehensive and analytical quantification of uncertainty, and because it can be made to interpolate observed simulations.

Typically, these simulators are deterministic (running the simulator with the exact same inputs multiple times gives the same outputs) and often arise from physics, engineering, or other similar fields. Stochastic simulators are also developed (running the simulator with the exact same inputs multiple times gives different outputs), often arising in fields such as epidemiology, ecology, and economics, as a result of the complexity of the natural or human world.

With any statistical model, from standard linear regression models used to estimate the effect of particular treatments to convolutional neural networks used to perform facial recognition, it is important to conduct tests to check whether the model is sufficiently accurate for the the problem at hand. Checks for deterministic emulators have been explored \citep{bastos2009diagnostics}, but we find these are not always suitable for stochastic emulators, and so this article discusses how to best validate a stochastic emulator.

\subsection{Emulation}

Formally, a Gaussian process emulator ($GP$) is a stochastic process where, for every  set of inputs $(\textbf{x}_1,\dots, \textbf{x}_N)$ and every finite $N$, the outputs $(y(\textbf{x}_1),\dots, y(\textbf{x}_N))$ are multivariate normally distributed with mean $m(\textbf{x}_1),\dots, m(\textbf{x}_N))$ and covariance matrix $K_N$ with entries $K(\textbf{x}_i, \textbf{x}_j)$ \citep{rasmussen2006gaussian}. Less formally, a mean function $m(\textbf{x})$ models global trends, and can be used to provide prior beliefs about the shape of the relationship; and a covariance function $K(\textbf{x},\textbf{x}')$ captures local structures, allowing for a flexible overall relationship. An additional diagonal term, $\sigma^2$, is often added to the covariance matrix because of numerical stability issues otherwise \citep{andrianakis2012effect}, although it has also been argued to provide improved predictive performance as well \citep{gramacy2012cases}.

We can write this as: 

\begin{equation}
\label{eq:homGP}
y(\textbf{x}) \sim GP(m(\textbf{x}), K(\textbf{x},\textbf{x}') + \sigma^2)
\end{equation}

Using a set of initial simulations $\textbf{y}$, with inputs $X$, the parameters within the Gaussian process emulator (i.e. the parameters within the chosen mean function and covariance function) can be learnt. Future simulations, with inputs $X^*$, can then be predicted using the following equation:

\begin{align}
\label{eq:homGP_pred}
y&(X*)\  |\  \textbf{y} \sim N\big(\mathcal{M}(X^*), \mathcal{V}(X^*)\big); \\
&\mathcal{M}(X^*) = m(X^*) + K(X^*, X)\big( K(X,X) + \sigma^2I \big) ^{-1} \big(\textbf{y} - m(X)\big), \nonumber \\
&\mathcal{V}(X^*) = K(X^*, X^*) + \sigma^2I - K(X^*, X)\big( K(X,X) + \sigma^2I \big) ^{-1}K(X, X^*) \nonumber
\end{align}

where $I$ is the identity matrix.

For stochastic simulators, the $\sigma^2$ term can be estimated and interpreted as the intrinsic noise of the simulator. Furthermore, because simulators are often complex, the assumption of constant noise (as made via a constant $\sigma^2$ term) can be a strong assumption, and can be relaxed (for example, \cite{boukouvalas2014optimal} consider parametric structures for the intrinsic noise). A common solution is to also model the intrinsic noise via another Gaussian process \cite{goldberg1998regression, kersting2007most, boukouvalas2009learning, binois2018practical}. With this setup, the simulator noise is assumed to be normally distributed, with the mean modelled as a Gaussian process and the simulator variance with a Gaussian process. This is then a very flexible class of models, allowing for complex shapes for both the mean and variance. Other models for stochastic simulators are possible, including those which do not assume normality \citep{plumlee2014building} and those which do not use Gaussian processes \citep{kilmer1999computing}, but the normal Gaussian Process approach is often capable, understandable, easily utilised, provides analytical uncertainty quantification, and is efficient with simulations. The assumptions made and the potential for alternative methods, however, makes diagnostic checks additionally important when normal Gaussian processes are used for stochastic emulation.

A capable R package exists for the coupled Gaussian process approach, \texttt{hetGP} \citep{binois2017hetgp}, using the implementation from \cite{binois2018practical}. An overview of stochastic emulation more generally is available in \cite{baker2020stochastic}. As an important comment, due to the nature noisy nature of simulations, more simulations are often needed to obtain a good emulator when compared to the deterministic case. \cite{loeppky2009choosing} recommends 10 simulations per input dimension for deterministic simulators, whereas in an investigation between different design considerations, \cite{binois2019replication} uses up to 25 times this number.

\subsection{Standard (Deterministic) Diagnostics}

As an introductory example, to showcase the importance of diagnostics for stochastic problems, and the current (deterministic) Gaussian process diagnostics, consider the toy stochastic simulator in Equation \ref{eq:toy_sim}

\begin{align}
\label{eq:toy_sim}
\begin{gathered}
y =  sin(16x) + cos(24x) + 8x + \epsilon;\\
\epsilon \sim N\big(0, (0.1 + 0.9x)^2\big)
\end{gathered}
\end{align}

\begin{figure}[!htb]
\centering
\begin{subfigure}{.5\textwidth}
  \centering
  \includegraphics[width=\linewidth]{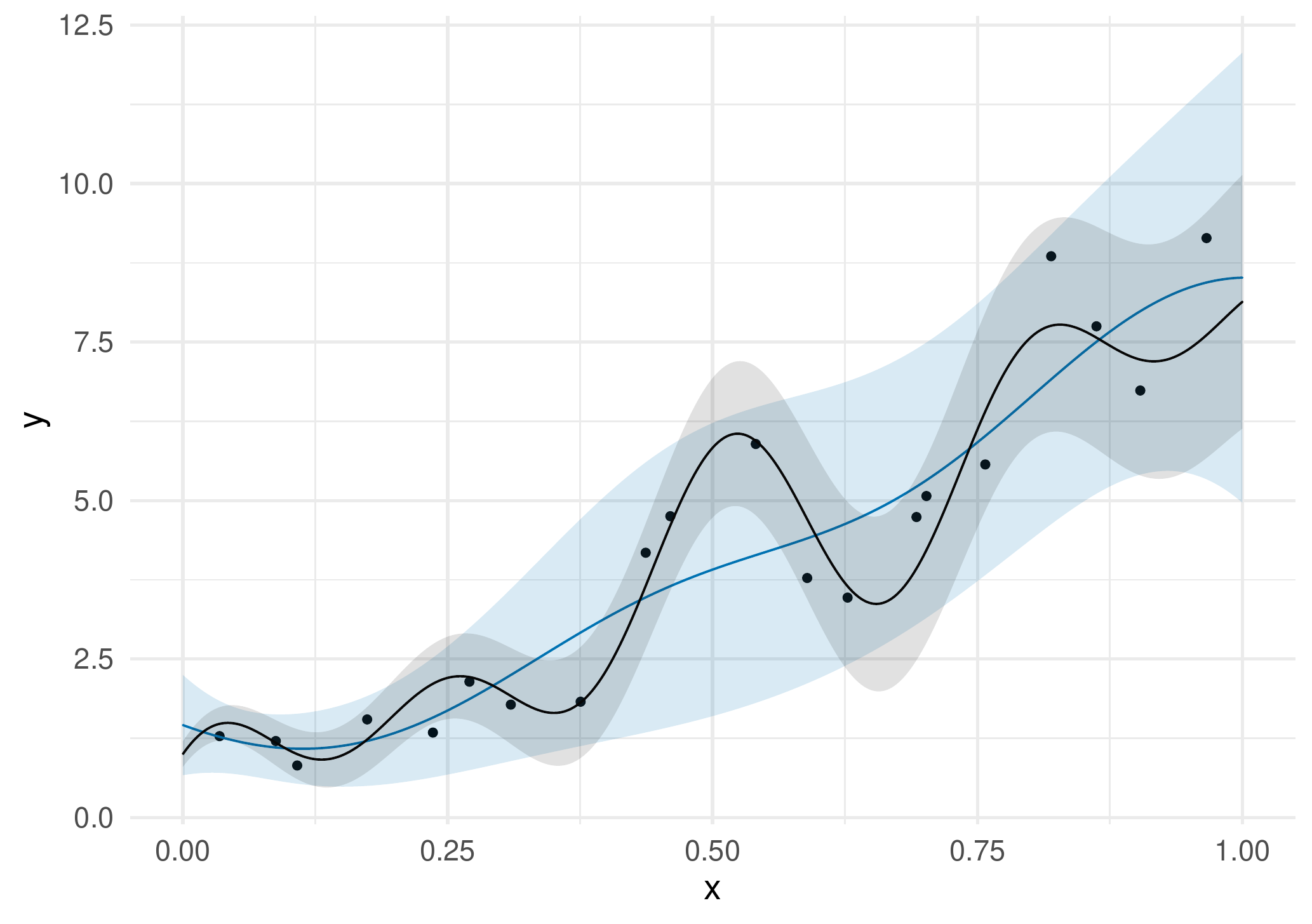}
\end{subfigure}%
\begin{subfigure}{.5\textwidth}
  \centering
  \includegraphics[width=\linewidth]{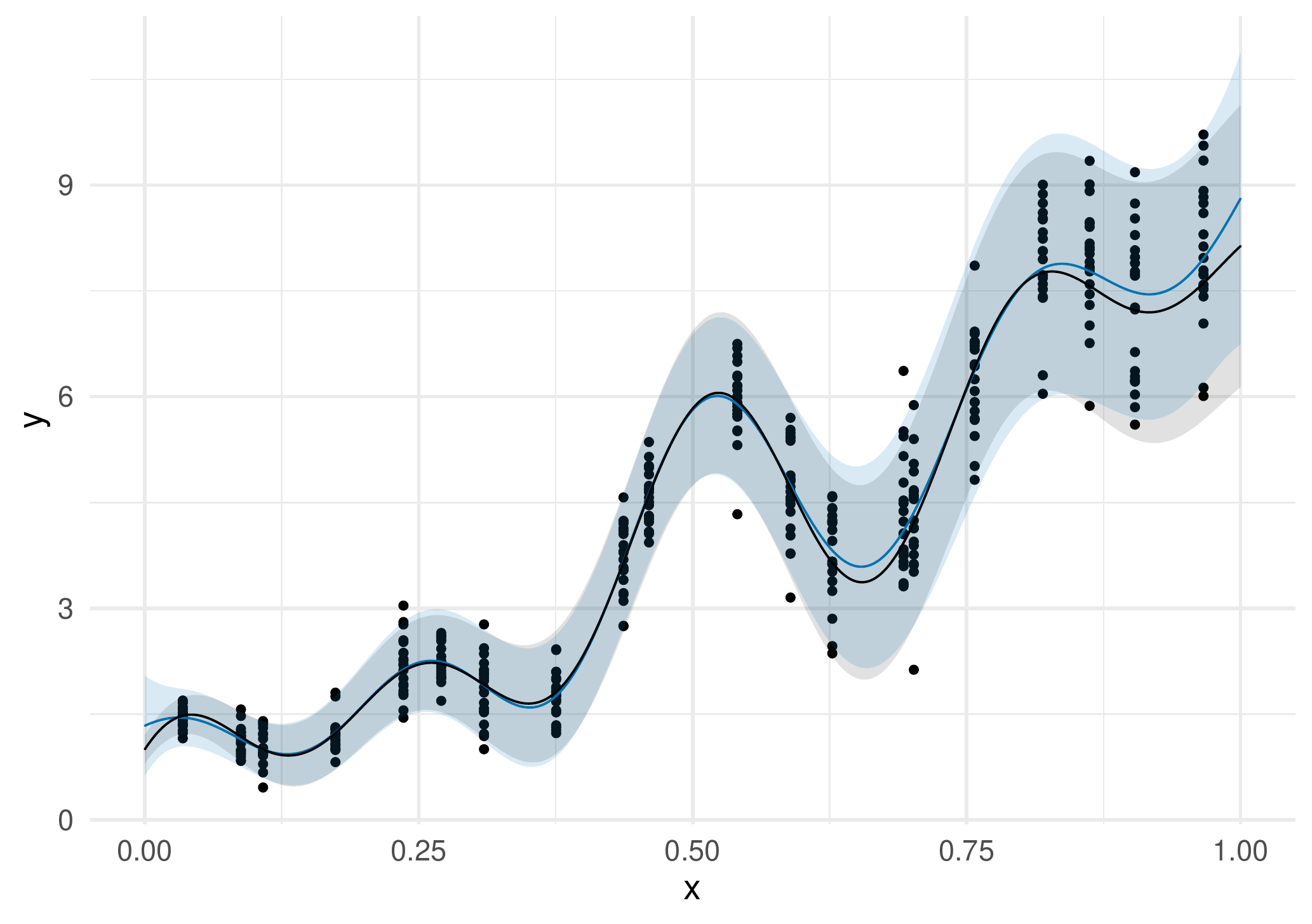}
\end{subfigure}
\caption{Two heteroscedastic Gaussian process emulators, fit to two different datasets, both from the same stochastic simulator. Blue represents the mean and 2 standard deviation intervals from the emulator. Black represents the truth.}
\label{fig:Initial_emuls}
\end{figure}

Figure \ref{fig:Initial_emuls} shows two sets of simulated data for this simulator, and their respective emulator fits using the \texttt{hetGP} package.

\begin{figure}[!htb]
\centering
\includegraphics[width=0.6\linewidth]{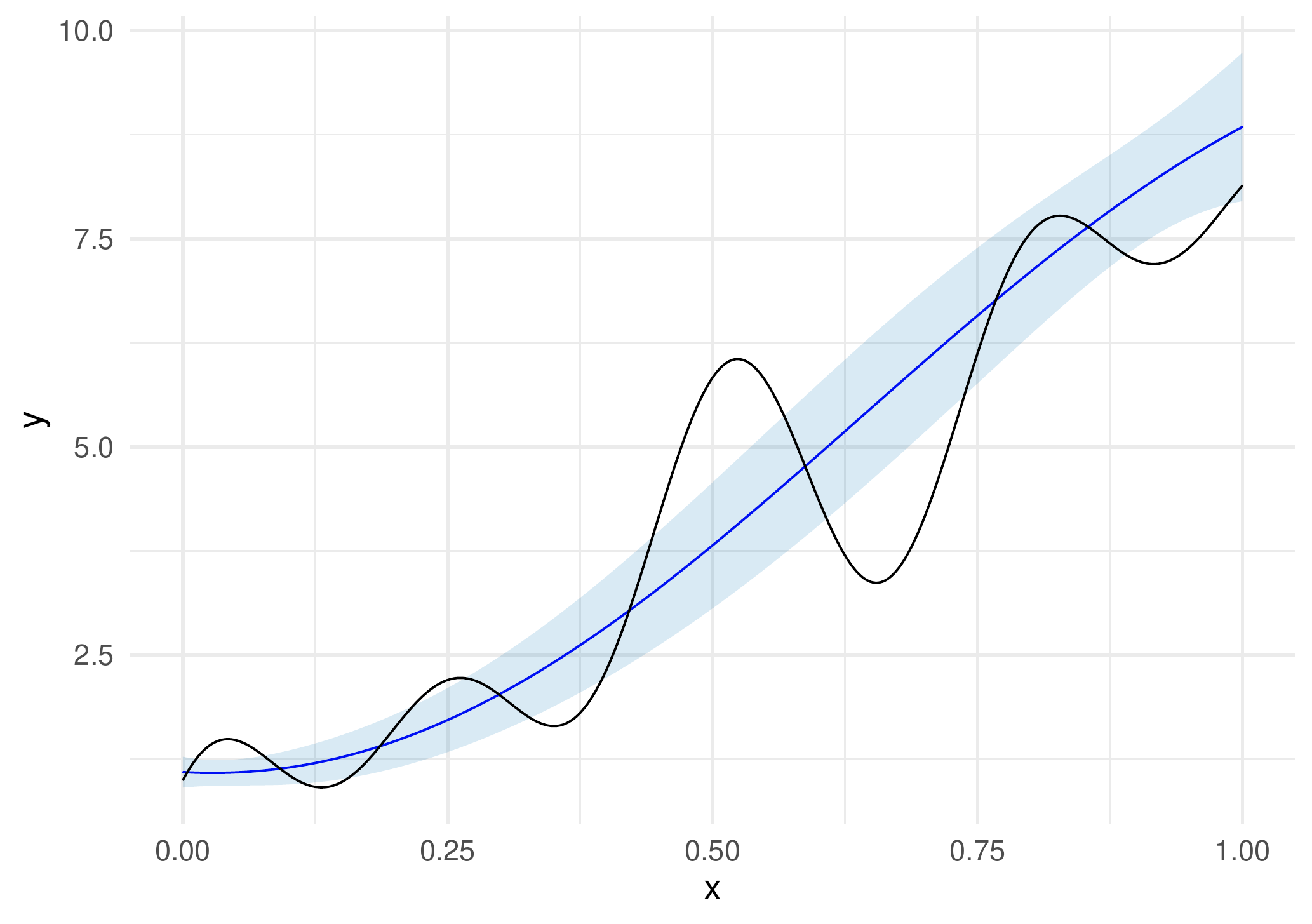}
\caption{The mean and 2 standard deviation prediction intervals for the mean from the `bad' emulator from the left of Figure \ref{fig:Initial_emuls} (in blue). Also superimposed is the true mean of the simulator (in black).}
\label{fig:initial_mean}
\end{figure}

Also shown is the truth. It should be immediately apparent that the emulator on the left is substantially worse than the emulator on the right. The emulator on the right appears to be a fairly good substitute for the simulator, accurately capturing the mean and variance. The emulator on the left however, fails to capture any of the nuance in the mean, and generally overestimates the variance. In fact, as figure \ref{fig:initial_mean} shows, the uncertainty intervals around the mean are also inaccurate, as the true mean is often outside of the emulator's $95\%$ prediction intervals for the mean.

This means that, not only is the emulator on the left less useful than the emulator on the right, it is also inaccurate. This type of emulator flaw, where an overestimated intrinsic variance leads to a poor estimation of the mean, is not a new discovery, and it has already been documented in the deterministic emulation literature where the inclusion of $\sigma^2$ (for numerical reasons) can cause this issue. \citep{andrianakis2012effect}.

Whilst details depend on the specific application and goal, using the emulator on the left could lead to incorrect conclusions, especially if the mean-variance structure is important. It can therefore be important to identify flaws like this.

In practice, identifying flaws can be difficult. Problems can often be of higher dimension, and access to the truth is rarely (if ever) available, and so simple plots like those above cannot be used to check the quality of an emulator. However, other checks can be done, often using a set of `validation data' (i.e. a set of data from the simulator, preferably not the same set used to fit the emulator), to assess the quality of an emulator. The more the emulator's predictions and the validation data match up, the better confidence one can have in the emulator.

\begin{figure}[!htb]
\centering
\includegraphics[width=\linewidth]{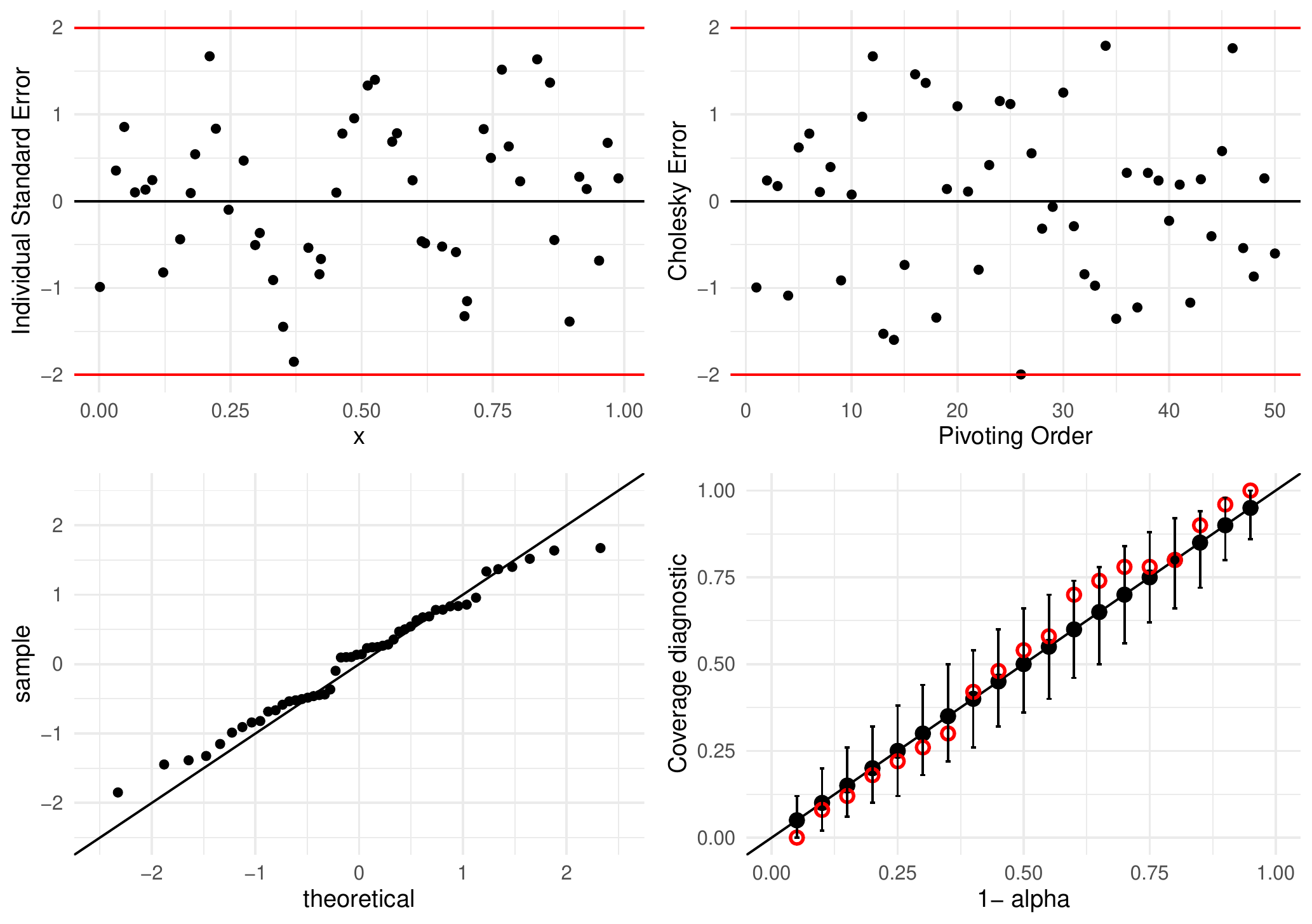}
\caption{Four diagnostic plots from \cite{bastos2009diagnostics} and \cite{al2018diagnostics} for the `bad' emulator. The top left is a plot of individual standardised errors against x; top right is a plot of the pivoted Cholesky errors against the pivoting order; bottom left is a QQ plot of the standardised errors; bottom right is a credible interval diagnostic plot.}
\label{fig:4_det_diag_bademul}
\end{figure}

\cite{bastos2009diagnostics} outlined and summarised a catalogue of diagnostic tools which could be used to diagnose problems in a deterministic Gaussian process emulator, which was later added to by \cite{al2018diagnostics}. After obtaining 50 validation data points from the simulator, the diagnostic plots in Figure \ref{fig:4_det_diag_bademul} can be obtained for the `bad' emulator. 

As a quick introduction on how to interpret these plots: the standardised errors and Cholesky errors should look like standard normally distributed noise about 0, the QQ plot should look roughly like the straight line, and the credible interval plot should also look like a straight line, with points falling within the intervals. More information about how to interpret these diagnostics can be found within the references. 

In general, these diagnostics do not flag anything wrong with the `bad' emulator. A keen eye might notice that none of the standardised errors have a larger absolute value than 2, which, with 50 validation points, is mild evidence towards the total emulator uncertainty being too large, but small standardised errors are often considered to be a good result. Additionally, one might also notice some mild structure in the individual standardised errors, but this can often be the case in good fitting 1D emulators as well.

The remainder of this article is organised as follows: Section \ref{sec:StoDiags} outlines the developed diagnostics, using toy examples to reveal their utility; Section \ref{sec:BuildingModel} then applies the diagnostics to a more practical example, showing how these diagnostics could be used in practice; and Section \ref{sec:Conc} provides concluding remarks.

\section{Stochastic Diagnostics}
\label{sec:StoDiags}

The previous section suggests that a component-wise diagnostic procedure could be a useful strategy; that is, specifically checking the mean and variance (and distribution) of a stochastic emulator separately. With this framing, diagnostics developed for deterministic emulators can be viewed as diagnostics for checking the overall accuracy of an emulator.

It is not possible to obtain values of the mean and variance from a simulator (at least, not often), and so directly validating the mean and variance is not possible. However, insights into the mean and variance can be obtained by replicating simulations. Replication enables the calculation of \emph{sample} means and \emph{sample} variances, which can then be compared with the predictions and estimates from an emulator. The diagnostic tools developed here provide techniques for using sample statistics to validate and diagnose a stochastic emulator.

We will first outline diagnostic methods for assessing the quality of the mean.

\subsection{Assessing the Mean}
\label{sec:MeanDiag}

If a random variable $Z$ is normally distributed with mean $\mu$ and variance $\sigma^2$ ($Z \sim N\big(\mu, \sigma^2\big)$), then the distribution of the sample mean $\bar{Z}$ of n samples is known to still be normal, but with a smaller variance  ($\bar{Z} \sim N\big(\mu, \sigma^2/n)\big)$. With this, we could implement the same diagnostics as before, but now for the sample means; comparing the observed values to the emulator's predictive distributions. However, this assumes we have knowledge of the true intrinsic variance values, which we do not. We could use the emulator's intrinsic variance estimates in the construction of mean diagnostics, but this would prevent truly independent, component-wise, checking of the emulator.

Without true values of the intrinsic variance, we can instead substitute in values of the observed sample variance from the replicated simulations at each point. With a limited sample size, the sample variance will not be particularly accurate, and so the distribution for the sample mean changes, is a standard result, and can be obtained from: 

\begin{equation} \label{eq:Samplemeantdist}
(\bar{y}_i(\textbf{x}_i^*) - \mathcal{M}(\textbf{x}_i^*))\frac{\sqrt{r_i^*}}{\hat{S}_i} \sim t_{r_i^*-1}.
\end{equation}

where $\bar{y}_i$ is the sample mean, and $\hat{S}^2_i$ is the observed sample variance.

While this distribution is is no longer normal\footnote{A t distribution is very similar to a normal distribution, especially if the degrees of freedom is large. With a limited number of replicated simulations, the degrees of freedom can be quite small, and so differences can emerge between the normal distribution and the t distribution.}, we can still use this to compare the observed values to the predictions from the emulator. standardised errors were a useful diagnostic tool, as they readily show how unexpected any given simulation is, assuming the emulator is correct. Since every standardised error should have the same distribution ($N\big(0,1\big)$), a consistent scale was obtained, making it easy to compare and identify trends and unexpected values. It is with this consistent scale in mind that we develop a new diagnostic. 

If the probability distribution can be obtained for a given observation, such as a sample means above, then the cumulative distribution is available, and thus one can calculate the probability of observing a value less than the realised observation. For a sample mean, that is: $P(\bar{y}(\textbf{x}_i^*) \leq \hat{\bar{y}}(\textbf{x}_i^*))$ where $\hat{\bar{y}}(\textbf{x}_i^*)$ is the observed sample mean. Unexpectedly large observed sample means will have a large value of $P(\bar{y}(\textbf{x}_i^*) \leq \hat{\bar{y}}(\textbf{x}_i^*))$ and unexpectedly small observed sample variances will have a small value of $P(\bar{y}(\textbf{x}_i^*) \leq \hat{\bar{y}}(\textbf{x}_i^*))$. It is with this as a foundation that we define the ``unexpectedness'' U as:

\begin{equation}
\label{eq:U}
U = 2 (0.5-P(Z \leq \hat{Z}))
\end{equation}

where $Z$ is some random quantity (e.g. the sample mean at a given coordinate) and $\hat{Z}$ is an observation for that random quantity. As $P(Z \leq \hat{Z}))$ is bound between 0 and 1, $U$ is always bound between 1 and -1. A large positive value for $U$ corresponds with an unexpectedly small value for the observation, and thus the emulator likely overestimates $Z$; and a large negative value implies an unexpectedly large value for $Z$ and thus the emulator likely underestimates $Z$.

Unexpectedness values can be interpreted in much the same way as individual standardised errors (albeit inverted). Any $U$ with absolute value greater than 0.95 is (roughly) the same as any standardised error with absolute value greater than 2, as the 2 standard deviation intervals match (roughly) with the $95\%$ prediction intervals, and the 0.95 unexpectedness intervals match exactly with the $95\%$ prediction intervals. Similarly, any $U$ with absolute value greater than 0.995 is very unlikely and corresponds to an absolute standard error of -2.8.
Differences between the two plots in the normal case are there, especially for the extremes, as the unexpectedness should be uniformly distributed between 0 and 1, and the standardised errors should look standard normally distributed about 0, but a lot of information is conveyed in the same way.

\begin{figure}[!htb]
\centering
\includegraphics[width=\linewidth]{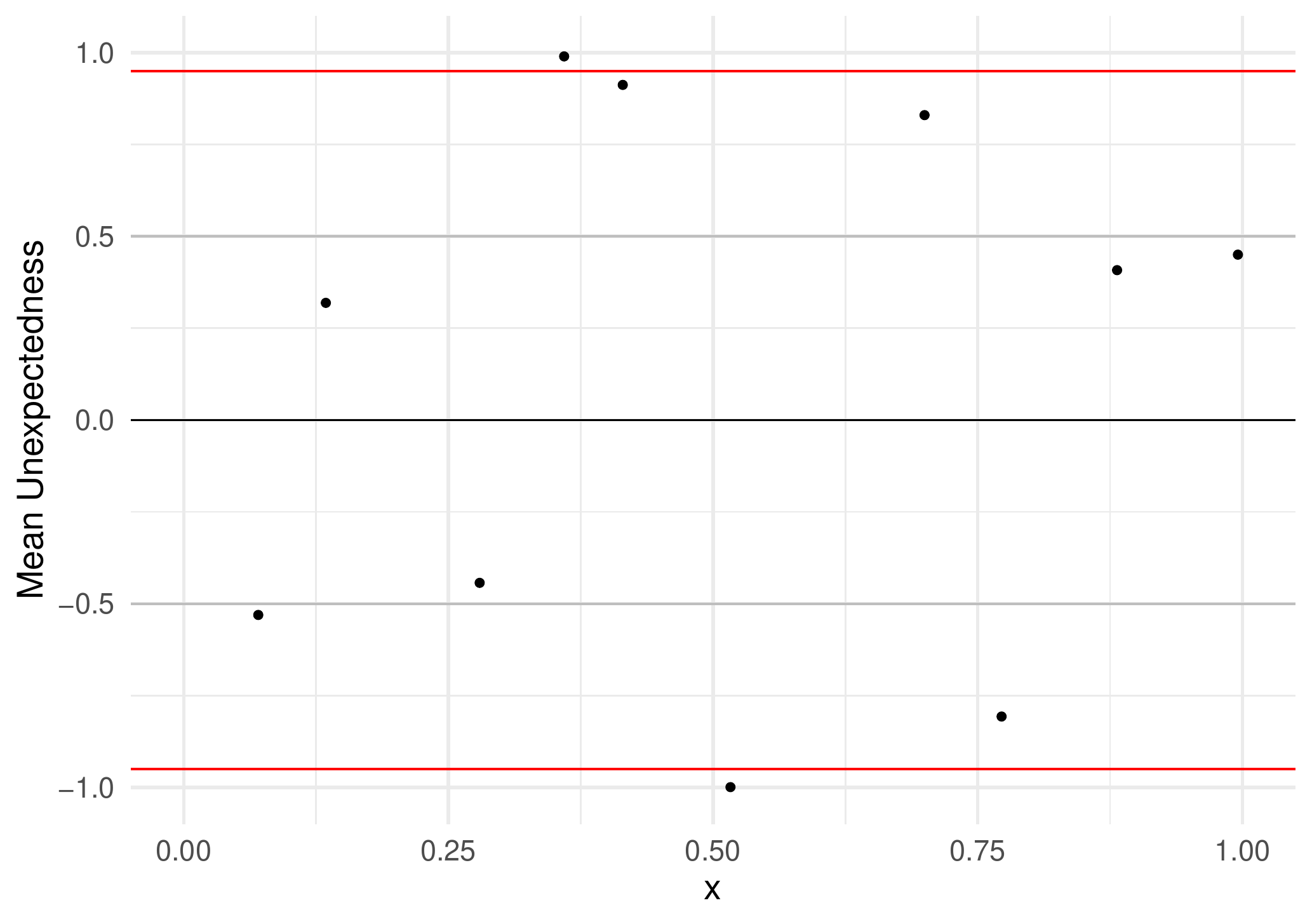}
\caption{Sample mean unexpectedness against $x$ for the `bad' emulator, using the sample variances to calculate the sampling distribution.}
\label{fig:Initial_MeantUs}
\end{figure}

Figure \ref{fig:Initial_MeantUs} plots the unexpectedness values against $x$ for the `bad' emulator. The cumulative sampling distribution is obtained by sampling predictive values of the mean $\mathcal{M}(\textbf{x}_i^*)$ from the emulator, and then propagating these through Equation \ref{eq:Samplemeantdist}. Note that if this analytical equation was not known, we could instead sample values from the emulator, sample multiple possible simulations using $S^2_i$, and then from these obtain many predictive values for the sample mean.

This uses the same total number of validation points as the previous diagnostics did in Section \ref{sec:Intro}, but now it is clear that there is a problem. Out of 10 sample means, two lie outside the 0.95 interval and one is smaller than -0.995, which is unlikely to happen by chance. Overall, these sample mean unexpectedness values provide substantial evidence that the emulator's mean is inaccurate. This is good for two reasons: firstly, they correctly identify a problem when previous diagnostics did not; secondly, they also provide information as to \emph{what} is wrong with the emulator. 

With knowledge of the truth (as we have access to Figure \ref{fig:Initial_MeantUs}), we can see that the two unexpectedness values with absolute value larger than 0.95 match up with locations where the true mean lies outside the uncertainty bound for the emulator's mean. However, there is no unexpectedness with absolute value larger than 0.95 around $x = 0.75$ which is another location where the truth and the emulator's mean do not match. This is because there is still residual noise in the sample means, and so not every problem has been identified, but a substantially better job has been done than before.

The unexpectedness has been designed in such a way that the plots are flipped compared to the standardised error plots: now large values of U imply the emulator's mean is too large, and small values imply it is too small (similarly for the sample variance unexpectedness later), which is perhaps a more natural interpretation. If the alternative is preferred, the scale can easily be flipped. The downside to the unexpectedness compared to the standardised errors is that extremely unexpected values are not quite as obvious - an unexpectedness value of 0.999 is much worse than 0.95, but appears similar when plotted. We can transform the unexpectedness in such a way to resolve this issue (using the inverse normal cumulative distribution function), but the unexpectedness alone has a fairly intuitive scale, and so we instead recommend direct interrogation of extreme values to check their value (perhaps even superimposing their value on plots).
For example, in Figure \ref{fig:Initial_MeantUs}, one of the unexpectedness values is as small as -0.9990, which is very unexpected (for reference, this would be similar to an standardised error of 3.29).

With this diagnostic, we have been able to invalidate the `bad' emulator (where previously we could not), we have learnt that the mean for this emulator has a poor fit, and we have developed a framework for assessing any component of an emulator.

\subsection{Assessing the Variance}
\label{sec:VarDiag}

With unexpectedness, diagnosing problems in the variance process can follow roughly the same procedure as with the mean: obtain the emulator predictive distribution for the sample variances and compare to the observed sample variances. For the sample variance, the sampling distribution is again known, and again not normal.
Given an estimate of the intrinsic variance $\sigma^2(\textbf{x}_i^*)$, the distribution for the sample variance $S^2$ can be obtained from:

\begin{equation} \label{eq:Samplevardist}
\frac{(r_i^*-1){{S}^2}(\textbf{x}_i^*)}{\sigma^2(\textbf{x}_i^*)} \sim \chi^2_{r_i^*-1}.
\end{equation}

\begin{figure}[!htb]
\centering
\begin{subfigure}{.5\textwidth}
  \centering
\includegraphics[width=\linewidth]{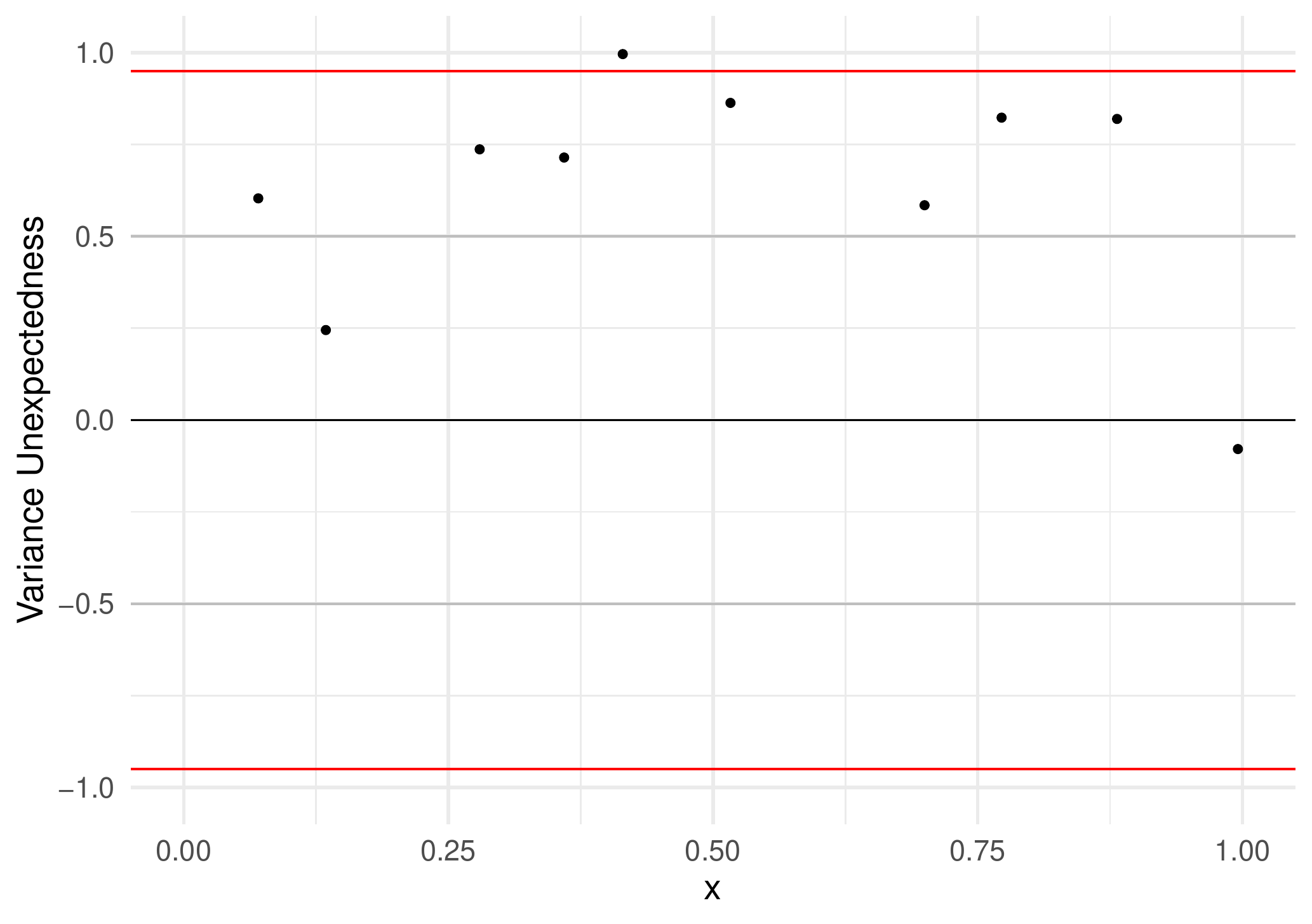}
\end{subfigure}%
\begin{subfigure}{.5\textwidth}
  \centering
\includegraphics[width=\linewidth]{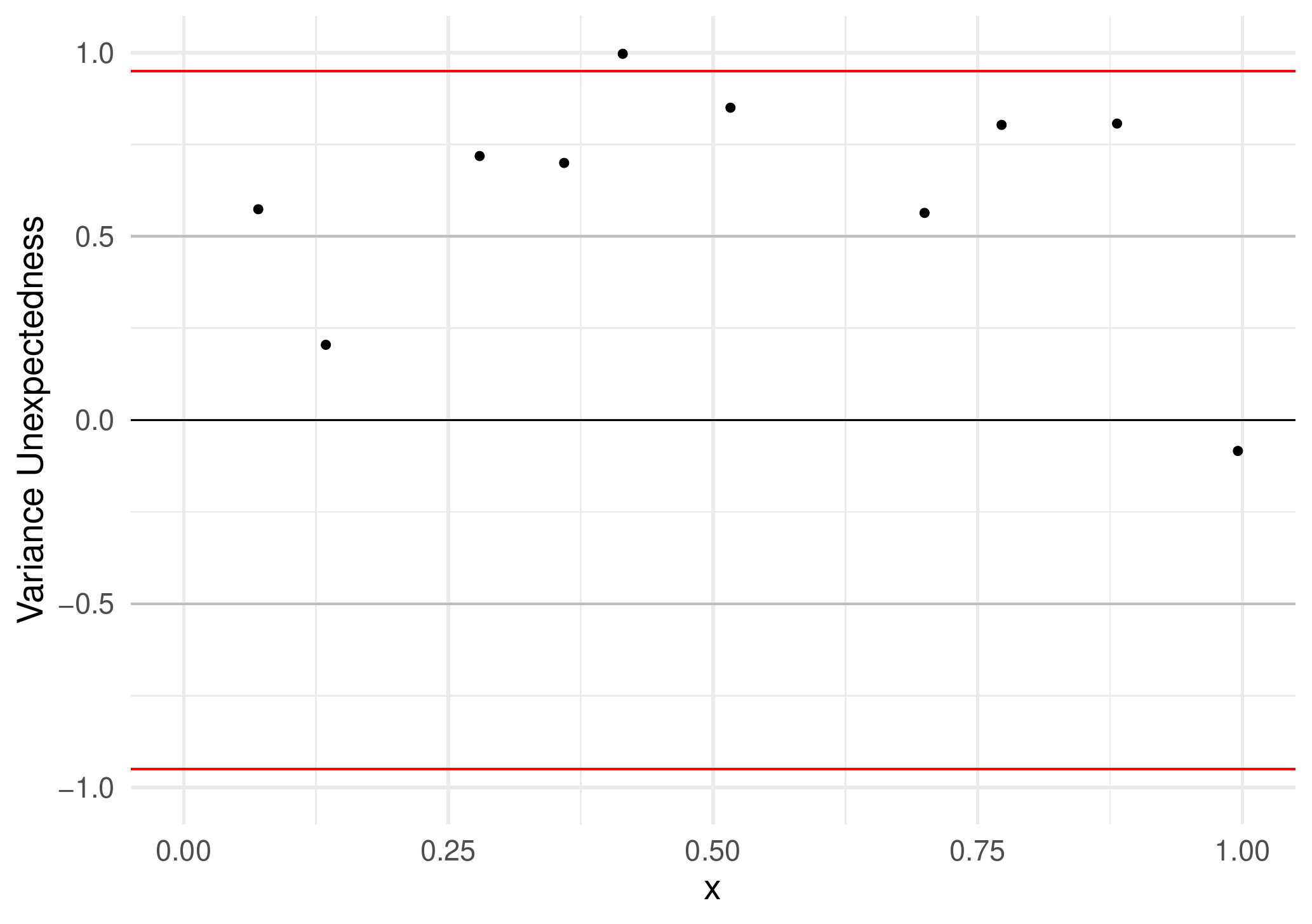}
\end{subfigure}
\caption{Sample variance unexpectedness against $x$ for the `bad' emulator. Left is the uncorrected version, right is the corrected version.}
\label{fig:Initial_VarU}
\end{figure}

A strength of unexpectedness is that it can be applied to almost any quantity as long as it is continuous and its distribution is known (or can be empirically estimated). As such, we can easily apply it to the sample variances, using Equation \ref{eq:Samplevardist}. The plot on the left in Figure \ref{fig:Initial_VarU} plots the sample variance unexpectedness against $x$ for the `bad' emulator.

From this there is evidence that the emulator overestimates the variance; nine out of ten of the unexpectedness values are positive, and one has a value larger than 0.95 (the value is  0.996, which is reasonably extreme). Together with the mean unexpectedness plot, these diagnostics reveal information that wasn't known before - the mean is poorly estimated, and the variance is systematically overestimated.

However, there is a problem with any variance process diagnostic for heteroscedastic Gaussian process emulators: most implementations do not feature any (or have reduced) epistemic uncertainty around the variance \citep{kersting2007most, boukouvalas2009learning, binois2018practical}. This is done for computational reasons, but it does raise an important philosophical question when it comes to validation: if there is no uncertainty around the variance estimates, does this mean every variance estimate is always wrong? Variances lie in the continuous domain, and so no estimate will ever be perfectly accurate. Without uncertainty, this means that, technically, every variance estimate is always wrong, and so we should always invalidate every heteroscedastic Gaussian process emulator, including the `good' one from Section \ref{sec:Intro} (and we will, given enough replication). 

We reject this notion, as it is not practical, in favour of a modification. We introduce a ``tolerance to error'' as a substitute for epistemic uncertainty around the variance. We do so by specifying a distribution that represents our tolerance to error for the variance estimates, and using that as if it were the epistemic uncertainty. We substitute the standard deviation estimates with the distribution $U(0.8\sigma, 1.2\sigma)$, which corresponds with a tolerance to any true standard deviation which is within $20\%$ of our estimate. Tolerance to error is a subjective notion, and so a practitioner might want to use a different distribution to quantify their tolerance to error. A uniform distribution was chosen here arbitrarily and implies indifference between a perfect variance estimate and an imperfect estimate within that is within the specified bounds. A triangular distribution tolerance-to-error would perhaps be more reasonable. 

The plot in the right of Figure \ref{fig:Initial_VarU} also presents the unexpectedness plot for the `bad' emulator including the tolerance to error. With only 5 replicates in the validation data, the tolerance to error is not necessary as the aleatoric sampling variation is still large, and so the results do not differ much. In either case, this is reassuring; the tolerance to error will safe-guard against heavily replicated validation datasets, improving the robustness of the diagnostics, but it does not seem to hinder validation in weakly replicated validation datasets. 

\subsection{Assessing the Normality Assumption}
\label{sec:NormalDiag}

With stochastic simulators, normality can be a strong assumption. Checking this assumption is therefore important if one is to have any confidence in the resulting emulator. A QQ plot of the standardised residuals is a standard method for testing the normality assumption, and is also recommended by \citep{bastos2009diagnostics}. The issue with QQ plots is that they only check for \emph{conditional} normality; i.e. whether the observed data is normally distributed around the mean, assuming the mean and variance estimates are correct. 

This means that a QQ plot should reveal problems if the simulator is not normal, \emph{or} if the mean and/or variance estimates are flawed. Similarly, because the mean (and variance) diagnostics rely on the normality assumption to some degree, these can wrongly reveal problems when instead the normality assumption is broken. This would be acceptable if the only purpose of diagnostics were to raise issues when they exist, but an important purpose of diagnostics is to also \emph{diagnose} what the problems are. As such, it would be helpful to have a diagnostic tool which assess the normality assumption in isolation, rather than assessing the conditional normality assumption.

Following a similar strategy as those used to assess the mean and variance, we make use of replicated simulations to check the normality assumption. If it is affordable, a single heavily replicated simulation could allow for visual assessment of the normal assumption via a histogram. However, this would be an inefficient use of simulation time, as this set of runs would not provide any information about the mean or variance across the input space, nor would it reveal issues if the normality assumption is broken but only in a different region of input space. As such, we would also like normality diagnostics which can use the same data as the other diagnostics, that can assess the normality assumption throughout the input space. 

Because normally distributed variables are characterised entirely by their first two moments (the mean and the variance), the later moments are the same for all normally distributed quantities, and these are what we will use to check the normality assumption. 

The third moment is the skewness, and represents the asymmetry of the distribution. A distribution with a positive skew will have a longer upper tail, and a distribution with a negative skew will have a longer lower tail. An important attribute of the normal distribution is that it is symmetric, and so for the normal distribution, the skewness should be 0. In much the same way as before, we cannot obtain values of the simulator skewness directly, only sample skewness from replicated simulations, and we can compare these to reference distributions to see how unexpected the sample values are.

The (standardised) sample skewness can be given by:
\begin{equation}
\label{eq:sampleskew}
\frac{1/n \sum (y_j - \bar{y})^3}{\hat{S}^3}
\end{equation}
where $\bar{y}$ is the sample mean and $S$ is the sample standard deviation. Taking the \emph{standardised} sample skewness helps to mitigate the impact of the variance. 

To obtain a reference distribution for this, we can generate data points using our emulator, obtain a generated value of the sample skewness, and repeat to obtain an empirical reference distribution for the sample skewness. We can then use this reference distribution and the observed sample skewness to calculate the unexpectedness using Equation \ref{eq:U}. Large values suggest the emulator needed to have assumed a distribution with a larger skew, and small values the opposite (in other words, large unexpectedness values suggest the simulator has negative skew, and small values suggest the simulator has positive skew).

We can repeat this exact procedure, but also for the fourth moment: the kurtosis. The kurtosis is a measure of how large the tails are. Kurtosis values are often compared to the normal distribution, which has a kurtosis of 3 (3 is often subtracted from the kurtosis, making the normal distribution the base case with an excess kurtosis of 0). Distributions with a kurtosis less than the normal distribution will have smaller tails and a lesser propensity to extreme values. Distributions with a kurtosis greater than the normal distribution will have larger tails and a greater propensity to extreme values. 

The (standardised) sample (excess) kurtosis can be given by:
\begin{equation}
\label{eq:samplekurt}
\frac{1/n \sum (y_j - \bar{y})^4}{\hat{S}^4} - 3
\end{equation}
where $\bar{y}$ is the sample mean and $S$ is the sample standard deviation.

And so the kurtosis unexpectedness values can be calculated along with the skewness unexpectedness values. Large kurtosis unexpectedness values suggest the emulator needed to have assumed a distribution with a larger tails, and small values the opposite (in other words, large unexpectedness values suggest the simulator has a negative excess kurtosis, and small values suggest the simulator has a positive excess kurtosis).
Together, these higher moments provide information about the simulator's distribution, and should allow for key non-normality structures to be noticed (such as heavy skewness or bimodality). Even higher moments could be used as well, but the fifth moment (and above) is less interpretable, less easily understood, and requires more data than the skewness and kurtosis .

These normality diagnostics suffer from the same problem as the uncorrected variance diagnostics - there is no epistemic uncertainty around the normality assumption. Whilst many real life processes appear roughly normal, they are rarely exactly normal. This then means that, technically, all diagnostics for normality \emph{should} discredit almost any quantity as non-normal. This won't happen often with the unexpectedness, however, as the number of replicates increases in a validation data set, the chance of correctly, but impractically, identifying non-normality increases. We can solve this problem in a similar way to the variance diagnostic correction: by adding a tolerance to error for the normality assumption.

Adding a tolerance to error to the normality assumption is more difficult, as there is no general, parametrised, distribution of which the Gaussian is a subset. However, there do exist extensions to the normal distribution which allow for skewness other than 0, and (excess) kurtosis other than 0. We can then add a tolerance to error using these extended distributions, and proceed as before.

For the skewness diagnostics, we make use of the skew normal distribution \citep{o1976bayes}. This is the normal distribution with an additional parameter $\alpha$, which affects the skew. A skew normal distribution with an $\alpha$ value of 0 exactly equals the normal distribution. The skewness of the skew normal is known and equals:
\begin{equation}
\begin{gathered}
\frac{4-\pi}{2} \frac{(\delta \sqrt{2/\pi})^3}{(1 - 2\delta^2/\pi)^{3/2}}\\
where \  \delta = \frac{\alpha}{\sqrt{1+\alpha^2}}
\end{gathered}
\end{equation}

We can then specify our tolerance to skewness error as $U(-0.5, 0.5)$, say, draw samples from this distribution, numerically invert them using the above equation, and then use the resulting $\alpha$ values to draw samples from the skew normal distribution (along with the emulator predictions for the mean and variance) to use in our calculation of the reference distribution for the sample skewness. This is then the equivalent of saying that we are ambivalent to true skewness values that are between -0.5 and 0.5, and so this is our tolerance to skewness error in the underlying intrinsic simulator variability. \footnote{$U(-0.5, 0.5)$ is subjective and depends on the specific application. In this case the values were chosen as values where the resulting theoretical distributions still (subjectively) appeared normal in histograms}

We can do similar for the kurtosis unexpectedness, using the generalised normal distribution \citep{nadarajah2005generalized}. The generalised normal distribution is similar to the skew normal distribution, in that it contains an additional parameter, $\beta$, which controls the kurtosis. A generalised normal distribution with a $\beta$ value of 2 exactly equals the normal distribution.
The (excess) kurtosis of the generalised normal is also known and equals:
\begin{equation}
\begin{gathered}
\frac{\Gamma(5/\beta)\Gamma(1/\beta)}{\Gamma(3/\beta)^2} - 3
\end{gathered}
\end{equation}

We can then follow the same procedure as for the skewness unexpectedness: specify our tolerance to kurtosis error (e.g. $U(-0.5, 0.5)$), draw samples from this distribution, numerically invert them using the above equation, and then use the resulting $\beta$ values to draw samples from the generalised normal distribution. This sampling procedure can then be used to obtain a reference distribution for the sample kurtosis, and thus can be used to obtain a kurtosis unexpectedness value.

We use both of these modifications in this article, although we do note that for the problems we examine, the modification has little effect.

As an example, consider the same toy simulator from Equation \ref{eq:toy_sim}, but where the intrinsic noise is instead gamma distributed (given again in Equation \ref{eq:Gammatoy}).

\begin{align}
\label{eq:Gammatoy}
\begin{gathered}
y =  sin(16x) + cos(24x) + 8x + \epsilon;\\
\epsilon \sim Gamma(1,1))
\end{gathered}
\end{align}

Following the same procedure as the `good' emulator from before (20 unique locations chosen via the same maximin Latin hypercube, each run 20 times, with a heteroscedastic GP fit, and 10 validation coordinates each replicated 5 times) should lead to diagnostics which flag issues, and they do:

\begin{figure}[!htb]
\centering
\begin{subfigure}{.5\textwidth}
  \centering
\includegraphics[width=\linewidth]{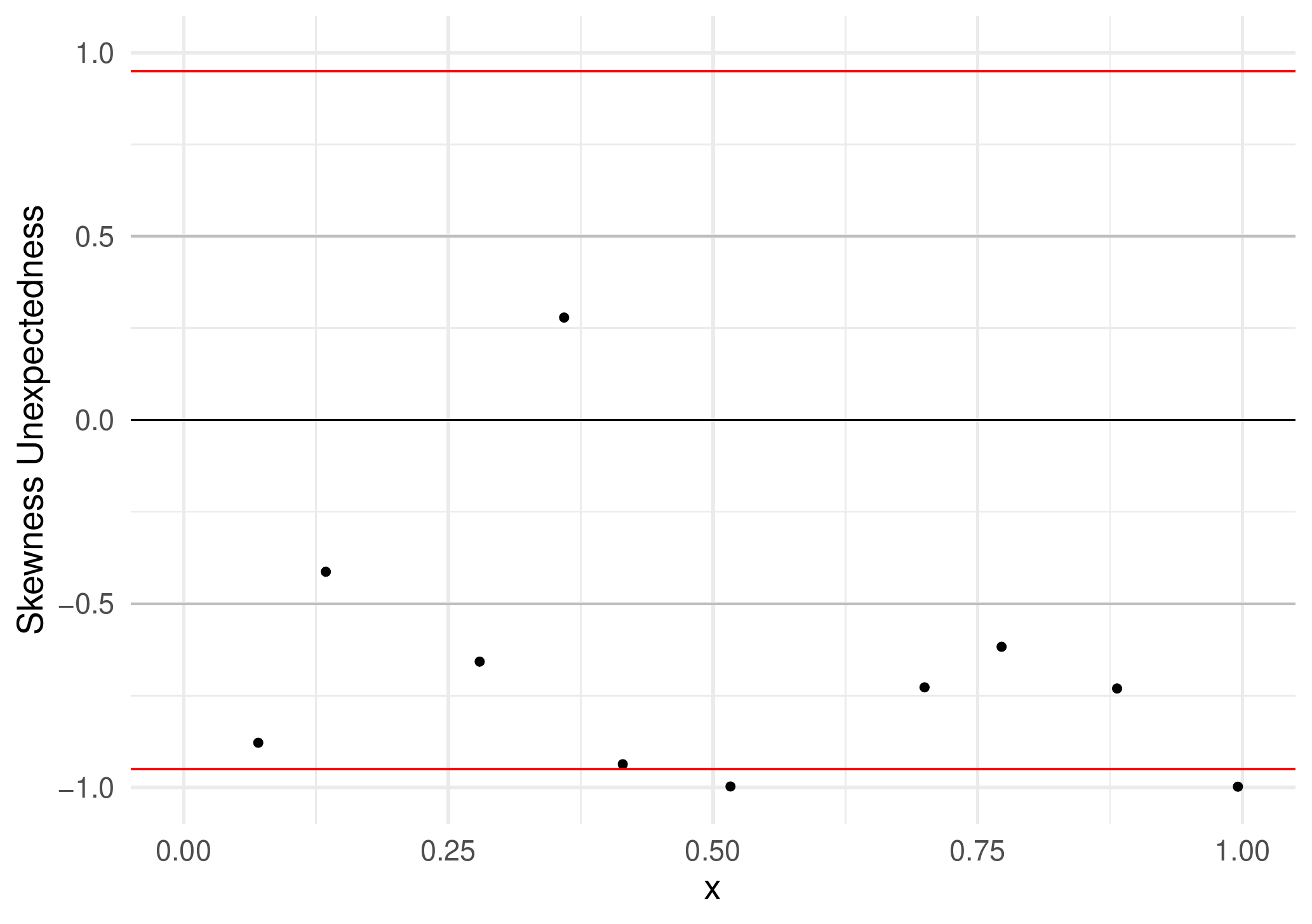}
\end{subfigure}%
\begin{subfigure}{.5\textwidth}
  \centering
\includegraphics[width=\linewidth]{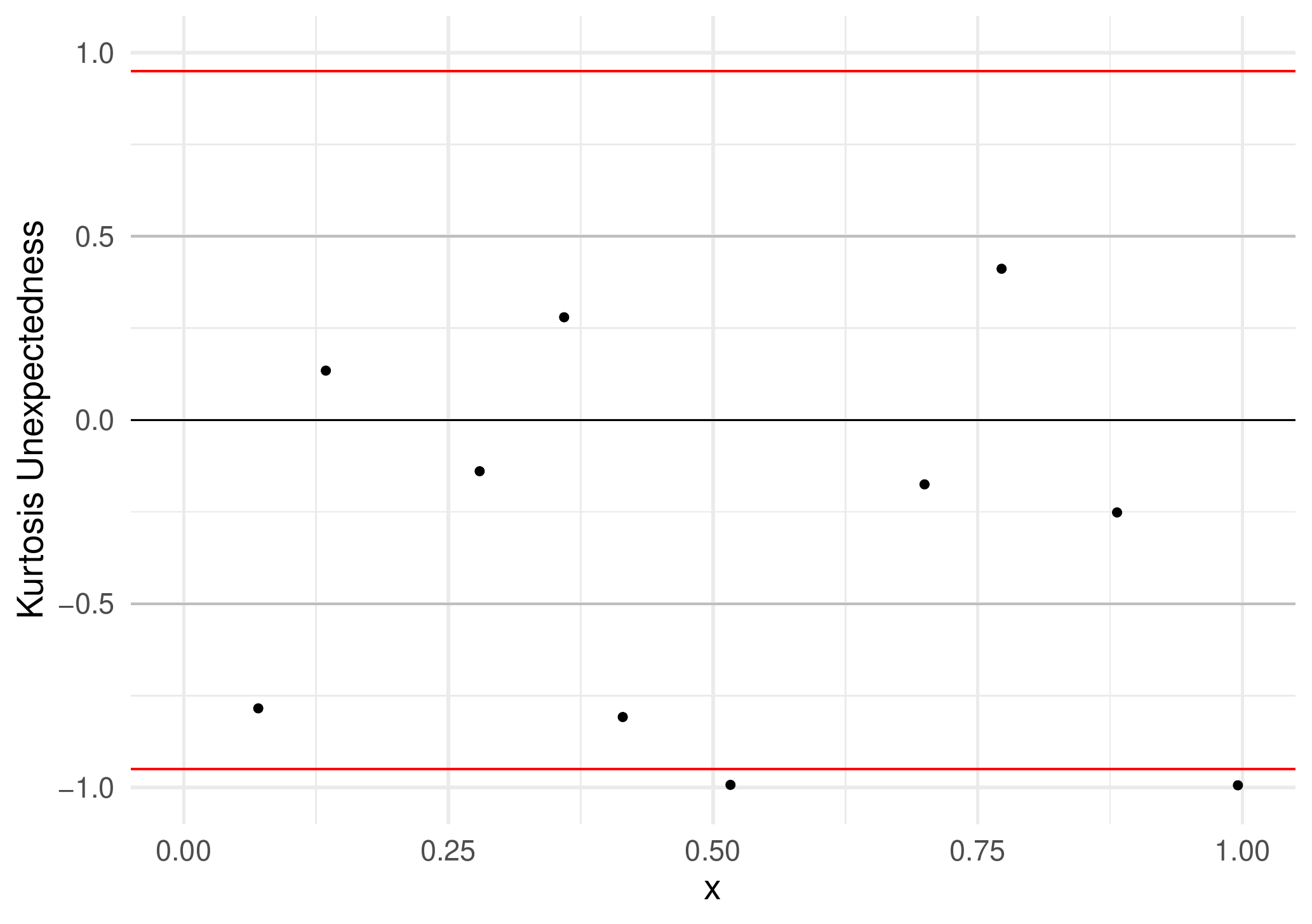}
\end{subfigure}
\caption{Skewness unexpectedness and kurtosis unexpectedness for the emulator of the gamma distributed toy simulator in Equation \ref{eq:Gammatoy}.}
\label{fig:Gamma_NormalityUs}
\end{figure}

The skewness unexpectedness plot in Figure \ref{fig:Gamma_NormalityUs} reveals the skewness of the emulator is too small (and so the skewness of the simulator is larger than the normal distribution. 9 out of 10 of the skewness unexpectedness values are less than 0, and 2 are less than -0.995. The kurtosis also reveals issues, with two unexpectedness values less than -0.95, suggesting a similar problem where the kurtosis of the simulator is larger than the normal distribution. Both of these statements are correct, and so we are then warned that the normality assumption may be insufficient.

This is reassuring, and suggests these normality diagnostics can identify problems. 

Overall, whilst these normality diagnostics are reasonably data efficient, they are more data hungry than the mean and variance diagnostics, as higher order moments are more difficult to precisely estimate. We need at least 3 runs per point to calculate the sample skewness, and at least 4 for the sample kurtosis. As such, if both of these diagnostics are desired, then 3 replicates are needed.

Note that other, standard, normality tests could be used here, repeated multiple times for each unique validation point, and the overall results used to assess the normality assumption (for example, the lilliefors test \citep{lilliefors1967kolmogorov} and the Shapiro-Wilk test\citep{shapiro1965analysis}). Care would have to be taken given the minimal number of replicates at each point, but a similar procedure as with the skewness and kurtosis unexpectedness could be followed. A downside of this is that these tests would not reveal much information about the non-normality of a simulator (if revealed), whereas the skewness and kurtosis diagnostics do reveal some information.

\section{Building Simulation Example}
\label{sec:BuildingModel}

\begin{figure}[!htb]
\includegraphics[width=\textwidth]{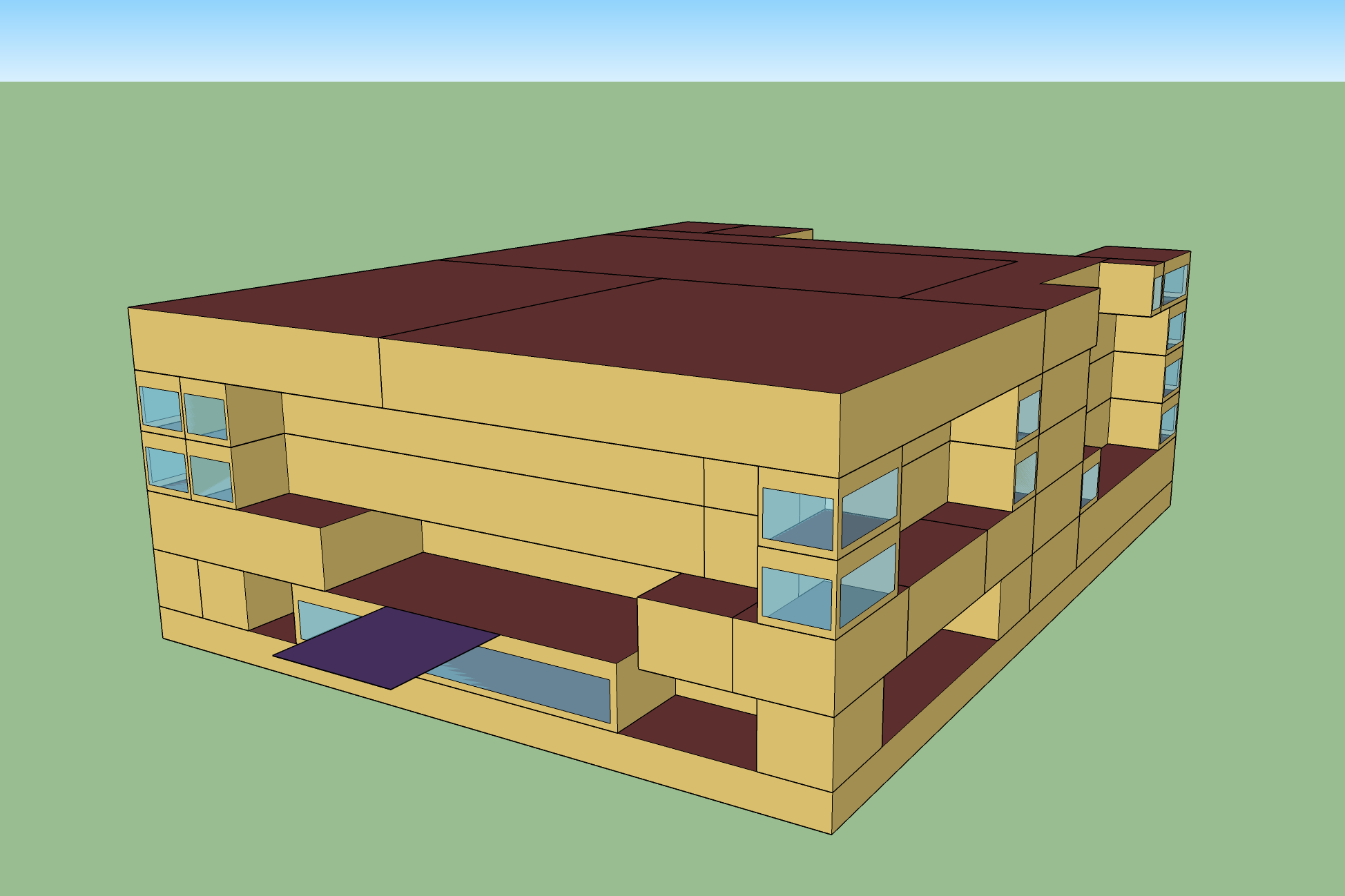}
\caption{The geometry of the modelled reference hospital building.}
\label{fig:Building}
\end{figure}

In this section, a building performance simulator will be used to showcase these diagnostic tools on a ``real'' example. A building performance simulator models output variables of a building, such as the internal temperature, given the physical shape and design of the building, and a description of the weather. EnergyPlus is one such building performance simulator \citep{crawley2000energy}. One use of this simulator is to estimate how much energy a planned building may use in a given time frame. We use a reference hospital design provided by the U.S. Department of Energy \citep{field2010using}. The building is given an `ideal loads' heating and cooling system, where the building will always be perfectly heated (or cooled) to the correct temperature with a subsequent increase in the buildings energy usage. The geometry of the building is presented in figure \ref{fig:Building}. 

For the weather input we use a very simplistic, stochastic, mechanism: we randomly stitch together weeks of historical observed weather, stratified according to season. The specific historical record we use is for Plymouth, using data from 1961 to 2016, and interpolating missing data. This makes the simulator stochastic. We then also consider as inputs to be adjusted (and the inputs to the emulator): the thickness of the wall concrete (varying from 0.05m to 0.75m), the thickness of wall insulation (varying from 0m to 0.4m), the roof insulation thickness (varying from 0m to 0.5m), the thickness of the ground concrete (0.05m to 0.75m) and the percentage size of the windows compared to the walls (25$\%$ to 75$\%$).
This is a simplified building model, and the choices made are not entirely realistic, but it can serve sufficiently well as an example.

To begin with, we perform 250 simulations; each with a unique input setting, with inputs chosen via a maximin Latin hypercube \citep{mckay2000comparison}. Our output of interest is the average energy usage of this building, which we model with a heteroscedastic GP using the \texttt{hetGP} package, with zero mean functions and squared exponential covariance functions. To then check the performance of this emulator, we apply the stochastic diagnostics outlined above. Validation points are chosen using another maximin Latin hypercube of size 50, with each point run 4 times.

\begin{figure}[!ht]
\includegraphics[width=\textwidth]{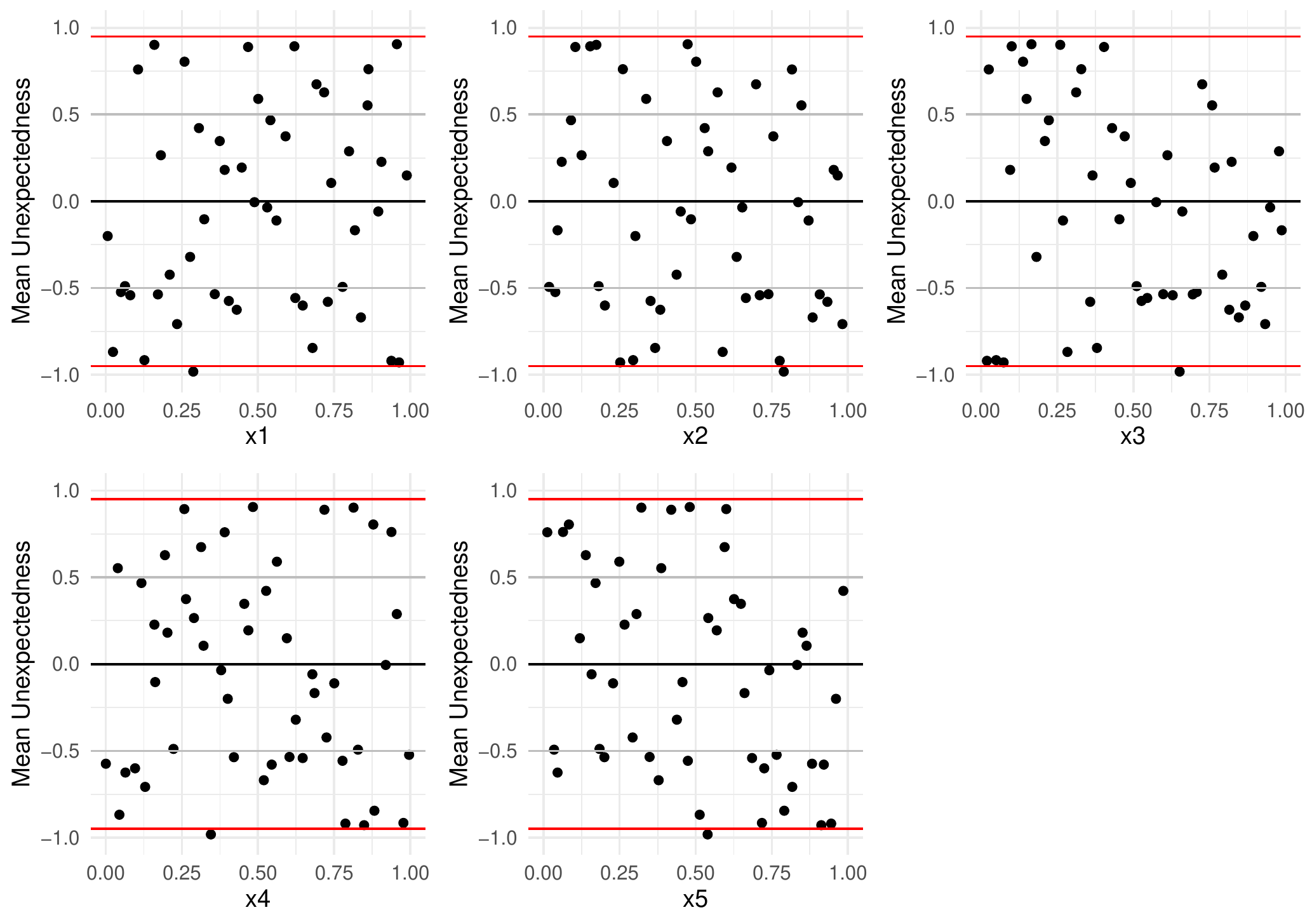}
\caption[Initial building mean diagnostics]{Sample mean unexpectedness for the emulator of the building simulator using 250 training points.}
\label{fig:Bad_building_MeantUs}
\end{figure}

\begin{figure}[!ht]
\includegraphics[width=\textwidth]{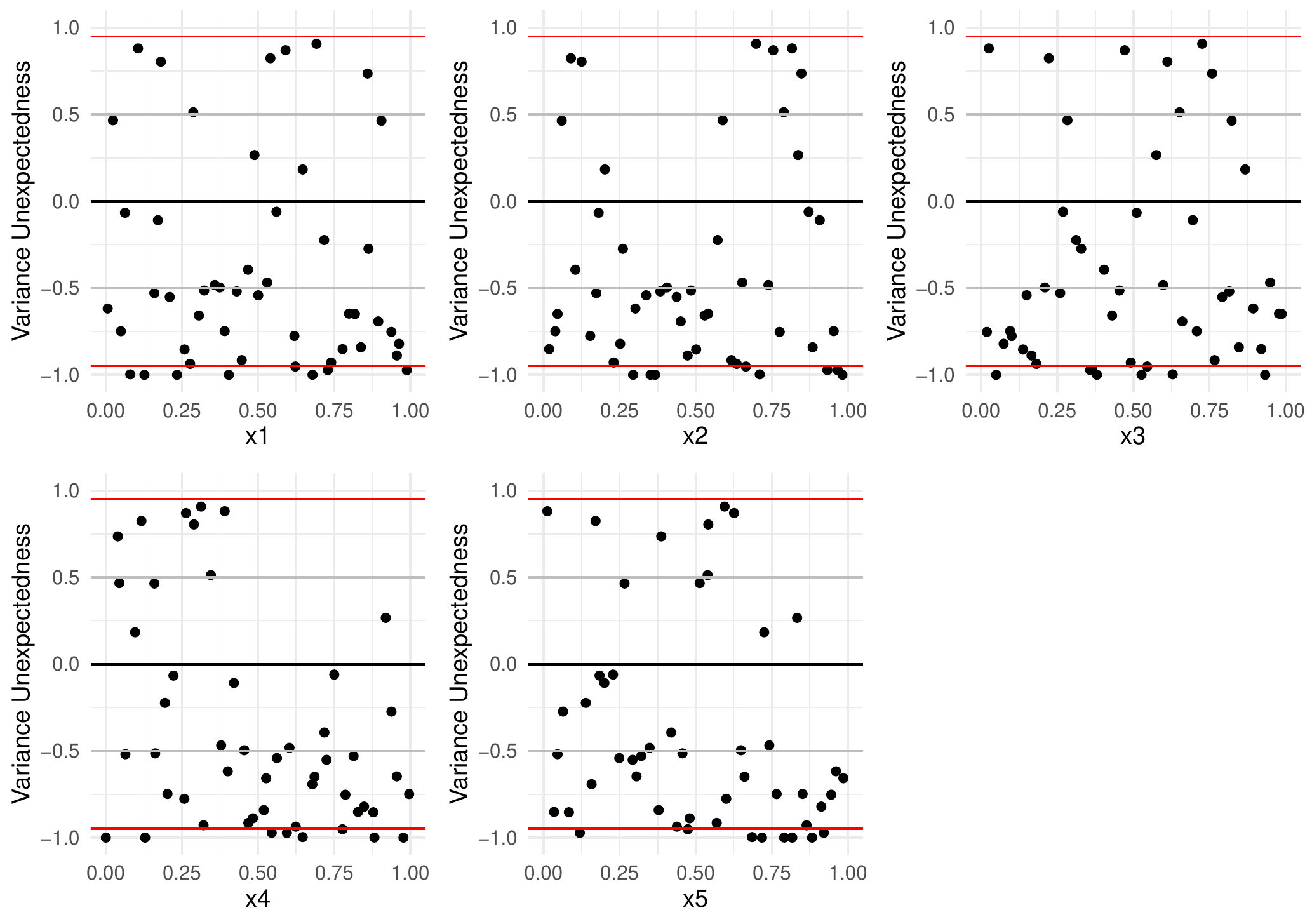}
\caption[Initial building variance diagnostics]{Sample variance unexpectedness for the emulator of the building simulator using 250 training points.}
\label{fig:Bad_building_VarUs}
\end{figure}

Figure \ref{fig:Bad_building_MeantUs} shows the sample mean unexpectedness for this building and \ref{fig:Bad_building_VarUs} shows the variance unexpectedness.

The mean diagnostics have only one unexpectedness value with absolute value larger than 0.95. With 50 total points, this is reasonable, suggesting that the mean is captured adequately. For the variance however, 8 are less than -0.95, 5 of which are less than -0.995. This suggests that, despite the reasonable mean predictions, the variance is underestimated. The plots don't seem to reveal any obvious patterns to this underestimation, suggesting it is a global issue.

Whether these flaws are a problem in practice will depend on the specific problem and objectives at hand. For example, the expectation of the stochastic ocean simulator in \cite{baker2020stochastic} corresponded with the underlying truth, and so in that case an good mean prediction but a poor variance prediction would be acceptable. Similarly, in \cite{baker2020future} the mean is the quantity of interest and so an underestimated variance would not be a problem. For this example however, we assume this flaw \emph{is} a problem. 

\begin{figure}[!ht]
\includegraphics[width=\textwidth]{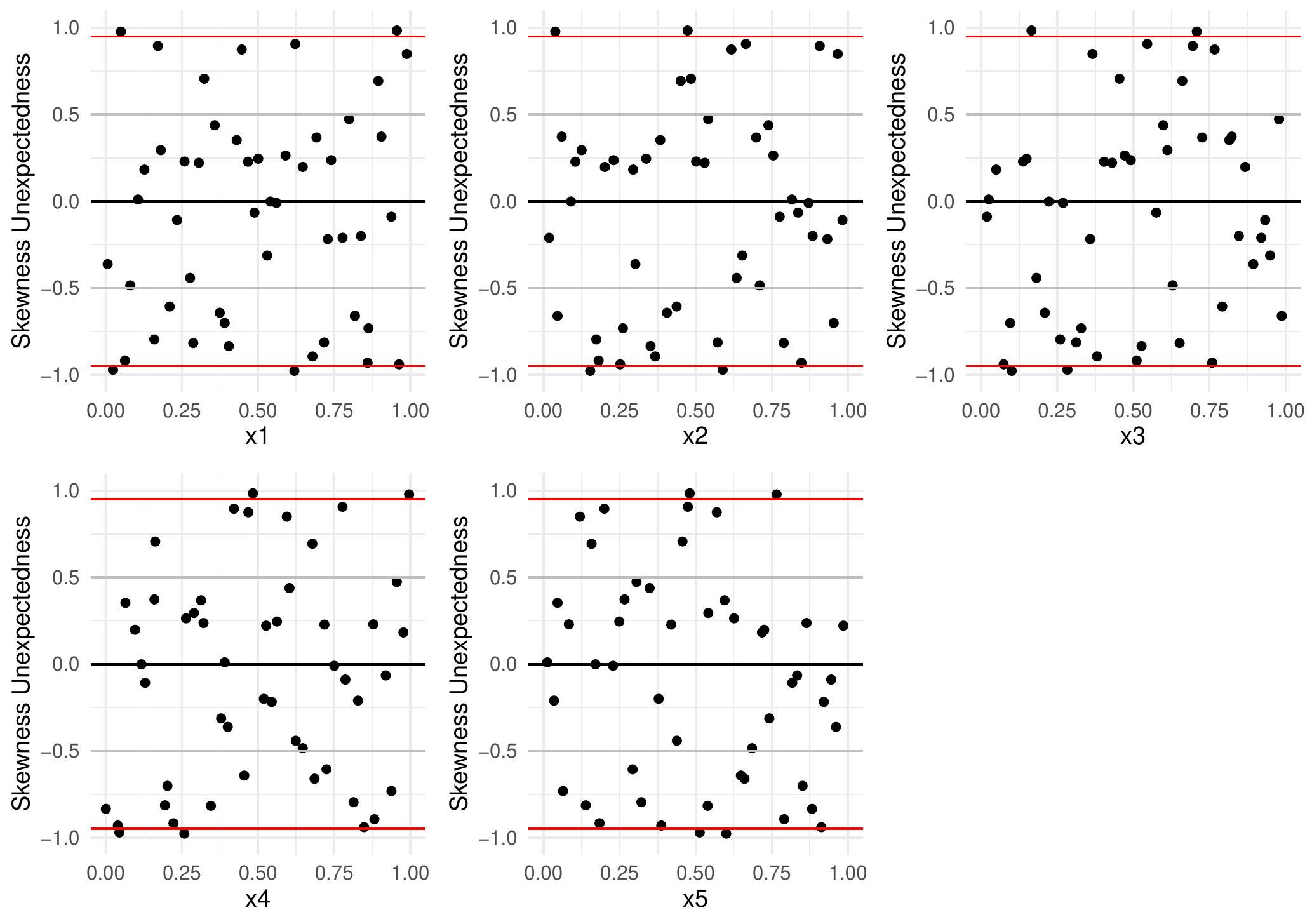}
\caption[Initial building skewness diagnostics]{Sample skewness unexpectedness for the emulator of the building simulator using 250 training points.}
\label{fig:Bad_building_SkewUs}
\end{figure}

\begin{figure}[!ht]
\includegraphics[width=\textwidth]{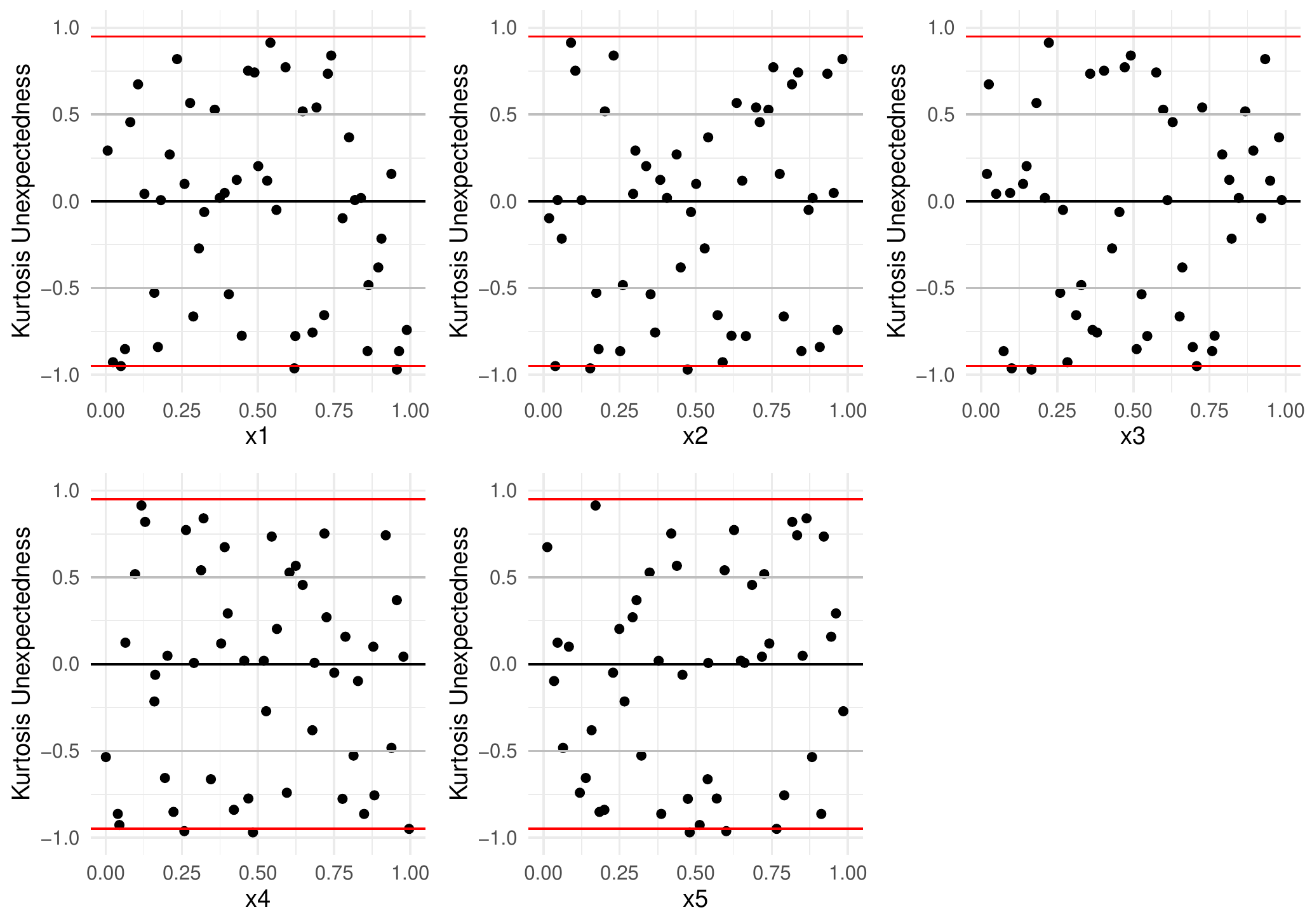}
\caption[Initial building kurtosis diagnostics]{Sample kurtosis unexpectedness for the emulator of the building simulator using 250 training points.}
\label{fig:Bad_building_KurtUs}
\end{figure}

To fix this emulator, a straightforward option would be to simulate further. Additional simulations should improve the emulator, and could provide improved variance predictions. This may not be realised however, if the underlying simulator variation is not normal. Non-normality could be the reason that the emulator is flawed, and it can also invalidate the diagnostics (potentially causing spurious results). We can use the normality diagnostics to check this, and therefore check whether the emulator variance issue is real, and whether the issue can be fixed with additional simulation.
Figure \ref{fig:Bad_building_SkewUs} and \ref{fig:Bad_building_KurtUs} show the skewness and kurtosis unexpectedness.

We can see from these plots, that the skewness unexpectedness does not reveal any glaring issues, with 4 skewness unexpectedness values with absolute value larger than 0.95, and 2 kurtosis unexpectedness values with absolute value larger than 0.95. This suggests (but does not guarantee) that the simulator is still normally distributed, and the emulator simply did not capture the variance process correctly.

To improve this emulator, we combine the validation data with the training data, and simulate an additional 250 points to make up the training data for a new emulator. Because our goal is to specifically improve the variance, replication may be beneficial here. We have some prior belief that the variance process will be simpler than the mean process (requiring fewer space-filling runs), and replicated simulations provide some insight into the variance at those specific locations. The merged validation data is already replicated 3 times, and our additional 200 runs are chosen to be the same as the initial 250 runs used to train the initial emulator. Together, this should provide enough information to capture the underlying mean and variance processes accurately.

\begin{figure}[!ht]
\includegraphics[width=\textwidth]{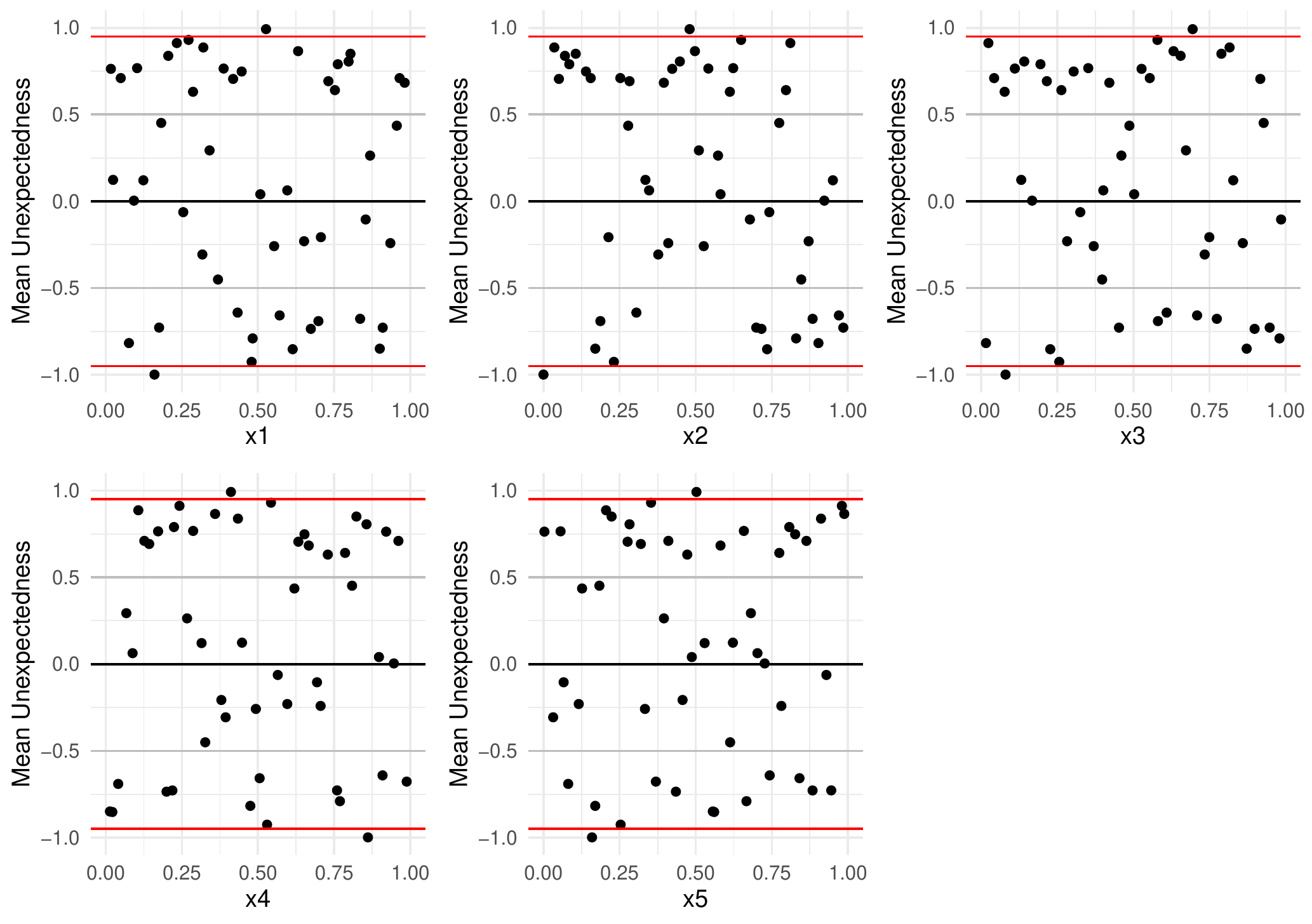}
\caption[Improved building mean diagnostics]{Sample mean unexpectedness for the emulator of the building simulator using the combined 700 training data points.}
\label{fig:Good_building_MeantUs}
\end{figure}

\begin{figure}[!ht]
\includegraphics[width=\textwidth]{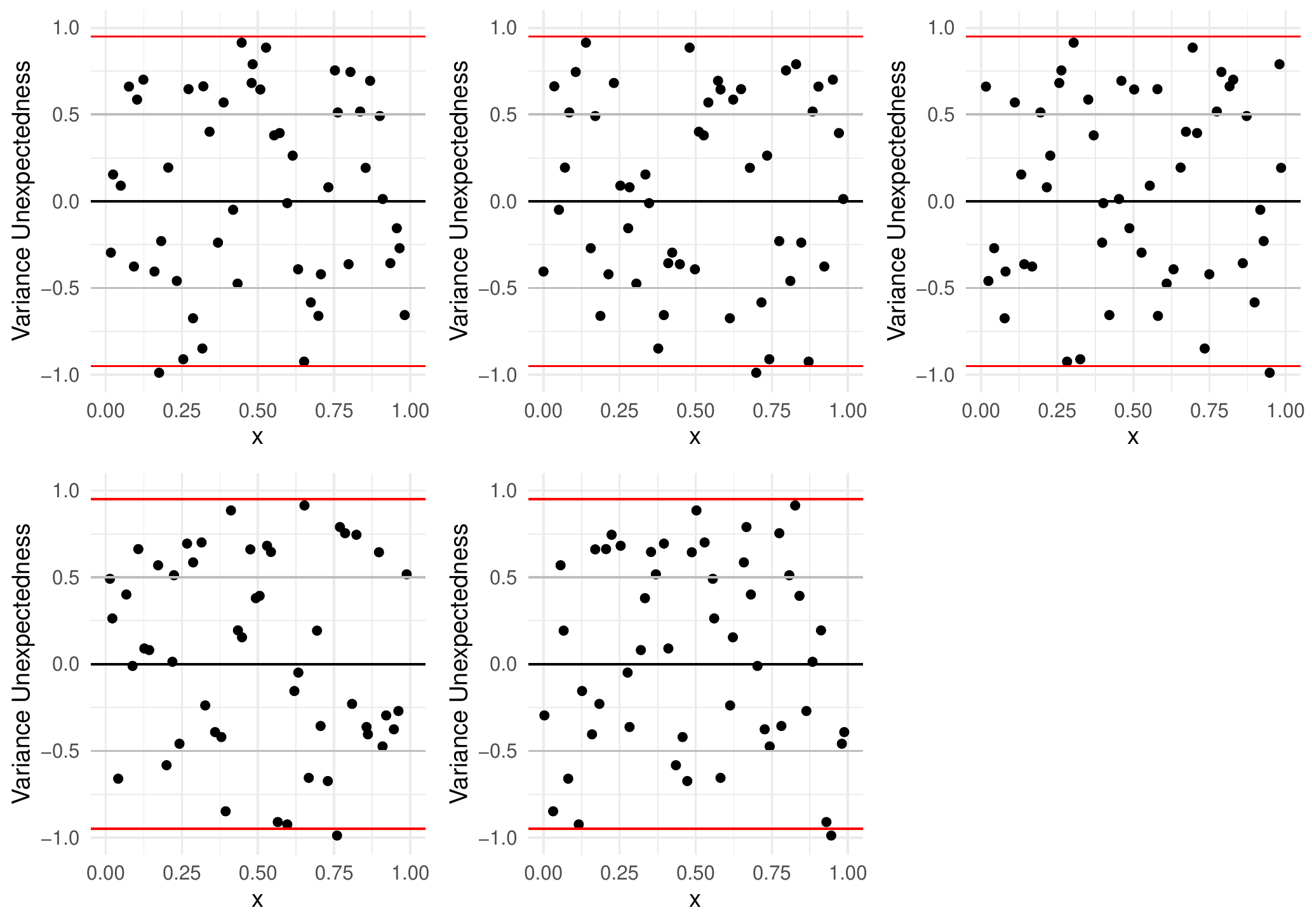}
\caption[Improved building variance diagnostics]{Sample variance unexpectedness for the emulator of the building simulator using the combined 700 training data points.}
\label{fig:Good_building_VarUs}
\end{figure}

To then check this new emulator, we obtain another 200 validation points, again by a Latin hypercube of size 50, each replicated 3 times. Figures \ref{fig:Good_building_MeantUs} and \ref{fig:Good_building_VarUs} show the same diagnostic plots as before.

These plots show a much improved emulator. Two mean unexpectedness values have absolute value greater than 0.95, and only one variance unexpectedness value has absolute value greater than 0.95, which is much more reasonable. There is however one mean unexpectedness value with a value of -0.9986, which is somewhat extreme, and would correspond to a standardised error of 3.19. It is not impossible for such a value to appear by chance, but it is less than ideal. Because the unexpected point is for a very small value of $x_2$ ($0.000155$), there could be a very local issue with the emulator's mean. In \cite{baker2020predicting}, this parameter ($x_2$; wall insulation thickness) was found to exhibit a steep change at low values, perhaps explaining this local emulator flaw. Additionally, for this example none of the 700 training data points were for an $x_2$ value as low as $0.000155$, and so it is not particularly surprising that the emulator may struggle here. We could improve this emulator (for example, by simulating more for low values of $x_2$), however, we decide that this emulator is acceptable, and would simply recommend some caution when dealing with very small values of $x_2$.

Normality diagnostics are not essential now, as we already concluded that the simulator is normally distributed. However, one of the skewness unexpectedness values has absolute value larger than 0.95, and none of the kurtosis unexpectedness values, both of which are reasonable.

Overall, this example has shown how the developed diagnostics can aide in the construction of a capable statistical model, and provide insights into the strengths and weaknesses of a given emulator.

\section{Conclusion}
\label{sec:Conc}

In this article we have discussed how stochastic emulators can be difficult to validate. Standard techniques are unable to identify key problems (at least without an abundance of data) and are unable to diagnose the reasons for a poorly fitting stochastic emulator. Our proposed solution involves using replicated validation simulator runs to individually assess the component assumptions and estimates of a stochastic emulator, providing greater insights into the performance of an emulator.

Previously developed emulator diagnostics can be adapted to apply to sample means; providing information regarding the emulator's mean predictions. We also developed additional diagnostics for the sample variance and the normality assumption. The developed diagnostic framework can also be applied to other quantities of interest, which could be useful for other, non-normal, emulators. 

A key benefit of separately checking the mean and variance functions is that, if we are more interested in the mean than the variance, then we can prioritise the importance of the mean diagnostics. Similarly for the variance. The component-wise validation can also provide additional information about how to improve the emulator, as seen with the illustrative example here.

For the examples in this article, we used an additional out-of-sample validation data set to use with these diagnostic tools, and this is the idealised validation procedure. We recommend 10 validation input locations per input dimension to fill space to some degree, hopefully avoiding this problem (following the same recommendation for fitting emulators in \citep{loeppky2009choosing}), but in higher dimensions more may be desirable. Additionally, we recommend 4 simulations per input location, enabling the use of all the diagnostics discussed here (but only 2 could be used if the simulation budget is especially low, facilitating only the mean and variance diagnostics, or leave-one-out could be performed instead if the training data includes replications). If the simulation budget is more limited, the same tools could be implemented using the training data via leave-one-out validation, assuming the training data included replication. Additionally, the normality diagnostics do not assess the emulator's parameter estimates or interpolative/extrapolative abilities, and therefore can safely be applied using training data (without needing to perform leave-one-out validation, again assuming the training data is replicated).

An issue with these diagnostics (and any practical diagnostic method) is that it is still be possible to validate an invalid emulator, or invalidate a valid emulator. For example; validation coordinates can, by chance, be chosen in areas where the emulator is indeed valid, but problems may exist in other areas of the input space. Similarly, although a validation coordinate may be correctly chosen in an area where the emulator has poor fit, the validation simulator runs for this coordinate may, simply by chance, provide sample means and sample variances that agree with the emulator's predictions. This is always a risk when dealing with random processes.

It is perhaps important to note that unexpectedness, the back-end to the developed diagnostic tools, is essentially a rescaling of p-values. We used a different term here to encourage slightly different usage, but ultimately the two are (almost) the same.

Additionally, it is important to note that no emulator will ever perfectly recreate a simulator. This raises frustrating questions about what the purpose of an emulator is, if all emulators are flawed. Ultimately the point of an emulator is for it to be fit for purpose, and as such, whether an emulator is valid depends on the experimental objectives. As such, hard rules should be (and have been) avoided. Instead it is recommended to treat these, and other diagnostics, simply as information to advise on the decision to `validate' or `invalidate' an emulator; and to provide specific details about how an emulator misrepresents the simulator.

Ultimately, it is difficult to effectively validate a stochastic emulator, at least without an abundance of data (where proper validation is most important). Nonetheless, the ideas presented here appear to be capable and economical, providing a descriptive set of tools for assessing the fit of a stochastic emulator.

\section*{Acknowledgements}
This work was funded in part by an Engineering and Physical Sciences Research Council studentship.

\bibliographystyle{apalike}
\bibliography{references}

\end{document}